# The optical absorption spectra of *V* centers in deformed MgO under pressure

Jie Zhang[1,2], Xian Long Wang[1], Kai Shuai Yang[1], Ya Cheng[1], Chuan Guo Zhang[1], Zhi Zeng[1,2,4*], Xian Ming Zhou[3], Hai Qing Lin[2]

[1]Key Laboratory of Materials Physics, Institute of Solid State Physics, Chinese Academy of Sciences, Hefei 230031, P. R. China

[2]Beijing Computational Science Research Center, Beijing 10094, P. R. China

[3]Institute of Fluid Physics, National Key Laboratory of Shock Wave and Detonation Physics, Mianyang 621900, China

[4]University of Science and Technology of China, Hefei 230026, China

*E-mail: zzeng@theory.issp.ac.cn

**Abstract**

The electronic structure and optical absorption spectrum of $V^0$ and $V^-$ center in MgO are investigated using first-principles calculations based on density functional theory. It is demonstrated that the configuration with distortion is energetically favorable in PBE0 functional. The six O atoms around Mg vacancy are non-equivalent due to the distortion. The defect states localized in the band gap are obtained by applying hybrid functional calculations, which is very important for optical properties. At ambient condition, the absorption peak at ~2.5 and ~2.6 eV is assigned to $V^0$ and $V^-$ center, respectively. With increasing pressure, the absorption peak shows blue shift. The trend is consistent with the experimental observation.

**Introduction**

   Magnesium oxide (MgO) is abundant in the earth lower-mantle and has attracted great interest for its wide range of applications such as medicine, cement, desiccant, catalysts and electronic devices. MgO is stable in a very large pressure range of 0-227 GPa[1]. Due to its simple structure and stability, MgO is used as calibration standard in static high pressure experiment. Besides, pure MgO is also a potential optical-window material in dynamic high pressure experiment because of its characteristic of transparence with wide band-gap. The properties of materials are related to macrostructures such as point defects which induce unique features in application. Cation vacancy ($V$ or color center) in bulk MgO, generated via deformation, is a kind of intrinsic point defects. Experimental results proposed that the optical-transparency loss of MgO is related to the defects [2]. As the intrinsic defects in MgO, Mg vacancy can trap one or two holes forming $V^-$ or $V^0$ center, and the holes localize on the oxygen site neighboring the Mg vacancy. However, it is difficult to obtain the correct charge on oxygen atoms near the Mg vacancy theoretically. The six oxygen atoms near the Mg vacancy is equivalent within the normal exchange-correlation interaction concern[3] with that the holes delocalize over the six neighbor oxygen atoms. Considering the non-equivalent O atoms neighboring Mg vacancy in the $V$-type centers, the geometry of original octahedral symmetry constructed by six O atoms is probably destroyed. As presented in the work of Droghetti *et al.*[4], the symmetry lowering around the vacancy site occurs when the initial ionic configuration is distorted around the vacancy site, and the Mg vacancy induces the defect level in the band gap which is closely related to the optical properties. The absorption peaks related to $V^0$ and $V^-$ center is located at 2.36 eV and 2.33 eV at room temperature in experimental study[5], respectively. However, the peaks of $V^0$ and $V^-$ center has not been reproduced and the properties under high pressure is still unclear. Therefore, knowledge of $V$-type centers is of special interest since it is a key to understanding optical properties of MgO. In the present work, the absorption peak of $V$-type center is well described if a certain amount of Hartree-Fork exchange is included, and the variation of the absorption peak with pressure is predicted.

In this paper, the first-principles calculations is used to research the structural and optical properties of MgO with *V*-type centers. The structure of MgO with $V^0$ and $V^-$ center is relaxed using different initial configurations. Then we compare the geometries and energies of the different initial configurations. In the configuration with distortion, the six O atoms surrounding Mg vacancy possess electrons with unequal values. Due to the effect of distortion, the local characteristic of *V*-type center is established based on hybrid density functional. The absorption peaks originated from $V^0$ and $V^-$ center are interpreted through theoretical simulations.

**Methods**

First-principles calculations are performed using the projector-augmented wave method [6,7] as implemented in the VASP code [8]. The 2×2×2 supercell containing 64 atoms is used for the perfect MgO. The defective cell is created by removing one Mg atom located at the center of the perfect cell. The Perdew-Burke-Ernzerhof (PBE) exchange correlation functional [9] is employed. Compared to experimental results, the band gap of MgO is severely underestimated with PBE calculation. While the hybrid-functional calculations produce a much better agreement with experiment. The hybrid-functional calculations are performed with the PBE0 functional which contains 25% of the exact exchange and 75% of the PBE exchange [10,11]. A 3×3×3 Gamma centered grid is generated for *k*-mesh. The lattice parameters and atomic positions are fully relaxed into their ground-state using a conjugate-gradient algorithm. The relaxation is stopped if forces and the full stress tensor are smaller than 0.01 eV/Å and 0.03 GPa, respectively. The energy cutoff of 520 eV is used, and the total energy is converged to $1.0 \times 10^{-6}$ eV in the electronic self-consistent loop. Spin polarization is included in the whole calculations.

**Results and discussion**

In the PBE0 calculations, the optimized lattice parameters and band gap are 4.195

Å and 7.5 eV, respectively, which are in good agreement with the experimental results (4.21 Å and 7.8 eV). In the process of moving one Mg atom, the neighbor atoms are perturbed and deviate original equilibrium positions, leaving one hole ($V^-$ center) or two holes ($V^0$ center) trapped on the O ions neighboring Mg vacancy. In order to obtain reliable results, it is the key to describing the Mg vacancy correctly before further research. By performing a complete structure relaxation, we obtain the ground state configuration. The local characteristic of $V^0$ and $V^-$ center is shown in Fig. 1 and Table I. The distance between Mg vacancy and O1 (O2) is smaller than the distance between Mg vacancy and the other four oxygen atoms at 0 GPa.

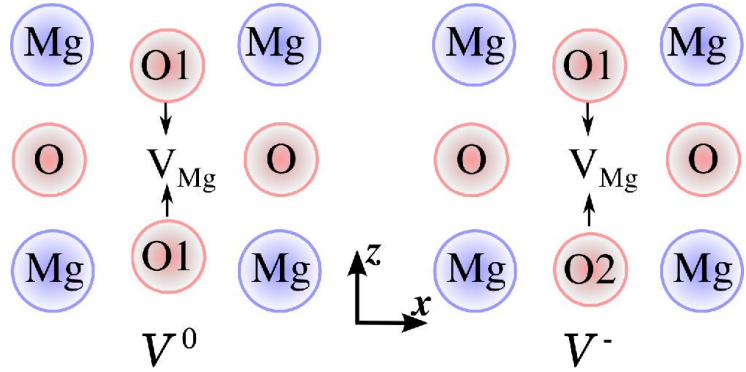

Figure 1. The ground configurations of $V^0$ and $V^-$ centers. The crystal structures are viewed along y-axis. The distortion of O1 and O2 atoms are different from the other four O atoms in $V^-$ center.

Table I. The optimized lattice parameters *a*, *b*, and *c*, and the distance from Mg vacancy to the six next-nearest-neighbor O atoms ($d_{V-O}$) are obtained in PBE0 functional for $V^0$-MgO and $V^-$-MgO (2×2×2 supercell) at 0 GPa. The experimental lattice constant is 4.21 Å. The number in brackets is the specific O atom with the distance ($d_{V-O}$).

|  | $V^0$-MgO | $V^-$-MgO |
| --- | --- | --- |
| *a*, *b*, *c* (Å) | a=b=8.41, c=8.37 | a=b=8.44, c=8.43 |
| $d_{V-O}$ (Å) | 2.21(4*O), 2.08(O1) | 2.24(4*O), 2.19(O2), 2.11(O1) |

To visualize the changes of MgO with *V*-type centers under high pressure, the structure is fully relaxed by the PBE0 functional and the structural parameters are listed in Table IS. As the results at 0 GPa, the oxygen atoms (O1 and O2) move inward with respect to the V-O distance of the other four O atoms around the vacancy. In order to determine the holes trapped on oxygen atoms in *V*-type centers, the charges on the six next-nearest-neighbor (NNN) O atoms of Mg vacancy are analyzed through Bader methods[12]. As shown in Fig. 2, the charges on the six O atoms exhibit small difference. The average charge on O1 is ~7.35$e$ in $V^0$ center, and the other four atoms possess more electrons than O1 atoms. While in $V^-$ center, the average charge on O1 atom is ~7.39$e$, and O2 atom and the other four O atoms possess almost ~8 electrons. In *V*-type center, O1 atom traps more holes than the other four O atoms, which means that *V*-type center is better described in PBE0 calculations. The symmetry broken is also important for the electronic structure of MgO with *V*-type center. In the PBE calculations, the defect state is merged with the valence band near the Fermi energy level. However, in hybrid functional calculations, the defect state mainly from O1-2$p$ orbitals lies in the middle of band gap (see Fig. 3).

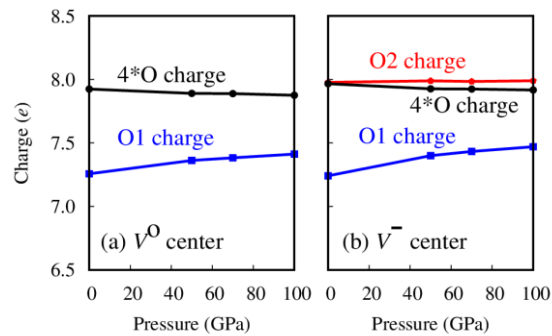

Figure 2. The charge of the six O atoms neighboring Mg vacancy. 4*O means the number of oxygen atoms with the same charge.

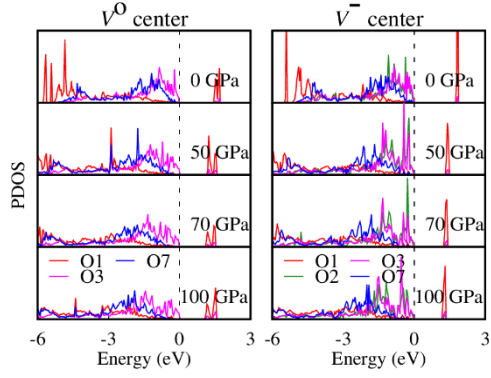

Figure 3. The density of states of $V^0$ and $V^-$ center at different pressures. The vertical dashed lines is the position of Fermi level. O3 represents the four equivalent O atoms neighboring the Mg vacancy, and O7 represents the farthest O atom from the Mg vacancy.

As a window material in shock wave experiment, the optical property is very important for interpreting the output signal. The optical properties are determined by the dielectric function $\varepsilon(\omega) = \varepsilon_1(\omega) + i\varepsilon_2(\omega)$, in which the real part $\varepsilon_1(\omega)$ and the imaginary part $\varepsilon_2(\omega)$ are connected by Kramers-Kronig relationship[13]. When we obtain the dielectric function from first-principles calculations, the absorption coefficient can be calculated by using the expression $\alpha(\omega) = \frac{\omega}{c}[2 \times (\sqrt{\varepsilon_1^2(\omega) + \varepsilon_2^2(\omega)} - \varepsilon_1(\omega))]^{1/2}$ [14]. The absorption peak of $V^0$ and $V^-$ center is shown in Fig. 4. At 0 GPa, the peak is located at 2.5 eV for $V^0$ center, which originate the electron transition from the four equivalent O atoms to O1 atom. Similarly, the peak at 2.6 eV of $V^-$ center is manly contributed by the electron transition from the four equivalent O atoms and O2 to O1 atom. With increasing pressure, the peak shows blue shift for both $V^0$ and $V^-$ center, which is consistent with the experimental result.

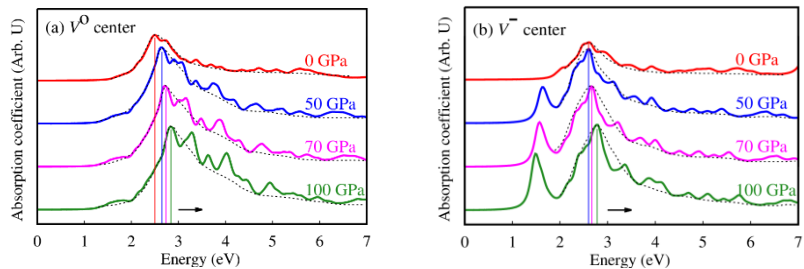

Figure 4. Pressure dependence of the absorption peak of $V^0$ and $V^-$ center in MgO. The dotted guide lines are plotted to mark the main peak.

**Conclusions**

The optical properties are calculated using hybrid density functional. The distortion around the Mg vacancy induces the non-uniform distribution of charge in the NNN O atoms. The defect state appears in the band gap. It results in the additional peaks located at ~2.5 eV and ~2.6 eV for $V^0$ and $V^-$ center, giving good agreement with experimental results.

**Acknowledgment**

This work was supported by the National Science Foundation of China under Grant Nos. 11504381, 11534012 and U1230202 (NSAF), the special Funds for Major State Basic Research Project of China (973) under Grant No. 2012CB933702. The calculations were partly performed in Center for Computational Science of CASHIPS, the ScGrid of Supercomputing Center and Computer Network Information Center of Chinese Academy of Sciences, and partly using Tianhe-2JK computing time award at the Beijing Computational Science Research Center (CSRC).

# Supplementary materials

**The optical absorption spectra of *V* centers in deformed MgO under pressure**

Jie Zhang[1,2], Xian Long Wang[1], Kai Shuai Yang[1], Ya Cheng[1], Chuan Guo Zhang[1], Zhi Zeng[1,2,4*], Xian Ming Zhou[3], Hai Qing Lin[2]

[1]Key Laboratory of Materials Physics, Institute of Solid State Physics, Chinese Academy of Sciences, Hefei 230031, P. R. China

[2]Beijing Computational Science Research Center, Beijing 10094, P. R. China

[3]Institute of Fluid Physics, National Key Laboratory of Shock Wave and Detonation Physics, Mianyang 621900, China

[4]University of Science and Technology of China, Hefei 230026, China

*E-mail: zzeng@theory.issp.ac.cn

Table IS. The optimized lattice parameters *a*, *b*, and *c*, and the distance from Mg vacancy to the six next-nearest-neighbor O atoms ($d_{V-O}$) at different pressures. The results are obtained in PBE0 functional. The number in brackets is the specific O atom corresponding to the distance ($d_{V-O}$).

| Pressure (GPa) | $V^0$-MgO | | $V^-$-MgO | |
|---|---|---|---|---|
| | Lattice parameters (Å) | $d_{V-O}$ (Å) | Lattice parameters (Å) | $d_{V-O}$ (Å) |
| 50 | *a*=*b*=7.86<br>*c*=7.83 | 2.00(4*O)<br>1.89 (O1) | *a*=*b*=7.88<br>*c*=7.87 | 2.03(4*O)<br>2.02(O2)<br>1.92 (O1) |
| 70 | *a*=*b*=7.72<br>*c*=7.70 | 1.95(4*O)<br>1.86(O1) | *a*=*b*=7.74<br>*c*=7.73 | 1.98(4*O)<br>1.97 (O2)<br>1.88(O1) |
| 100 | *a*=*b*=7.56<br>*c*=7.54 | 1.89(4*O)<br>1.80 (O1) | *a*=*b*=7.57<br>*c*=7.56 | 1.92(4*O)<br>1.91 (O2)<br>1.83 (O1) |